\documentclass{article}

\usepackage{caption}
\usepackage{subcaption}
\usepackage{PRIMEarxiv}
\usepackage[utf8]{inputenc} 
\usepackage[T1]{fontenc}    
\usepackage{hyperref}       
\usepackage{url}            
\usepackage{booktabs}       
\usepackage{amsfonts}       
\usepackage{amsmath}
\usepackage{amssymb}
\usepackage{nicefrac}       
\usepackage{microtype}      
\usepackage{lipsum}
\usepackage{fancyhdr}       
\usepackage{graphicx}       
\graphicspath{{media/}}     

\usepackage[dvipsnames,table]{xcolor} 
\usepackage{hyperref}
\hypersetup{
 unicode=false,          
 pdftoolbar=true,        
 pdfmenubar=true,        
 pdffitwindow=false,     
 pdfstartview={FitH},    
 pdftitle={My title},    
 pdfauthor={Author},     
 pdfsubject={Subject},   
 pdfcreator={Creator},   
 pdfproducer={Producer}, 
 pdfkeywords={keyword1, key2, key3}, 
 pdfnewwindow=true,      
 colorlinks=true,       
 linkcolor=purple,          
 citecolor=purple,        
 filecolor=purple,      
 urlcolor=purple           
}

\usepackage[noabbrev]{cleveref}

\usepackage{cite}

\title{FLRNet: A Deep Learning Method for Regressive Reconstruction of Flow Field  From Limited Sensor Measurements
}

\author{
  Phong C. H. Nguyen \thanks{\textbf{{Corresponding author}}: \texttt{phong.nguyenconghong@phenikaa-uni.edu.vn}} \\
  Faculty of Mechanical Engineering and Mechatronics \\
  Phenikaa University \\
  Hanoi 100000, Vietnam \\
  \texttt{phong.nguyenconghong@phenikaa-uni.edu.vn} \\
  \And
  Joseph B. Choi \\
  School of Data Science \\
  University of Virginia \\
  Charlottesville, VA 22903, United States\\
  \texttt{email@email} \\
  \And
  Quang-Trung Luu \\
  School of Electrical and Electronic Engineering\\
  Hanoi University of Science and Technology \\
  Hanoi 100000, Vietnam\\
  \texttt{trung.luuquang@hust.edu.vn} \\
}

\begin{document}
\maketitle

\begin{abstract}
Many applications in computational and experimental fluid mechanics require effective methods for reconstructing the flow fields from limited sensor data. However, this task remains a significant challenge because the measurement operator, which provides the punctual sensor measurement for a given state of the flow field, is often ill-conditioned and non-invertible. This issue impedes the feasibility of identifying the forward map, theoretically the inverse of the measurement operator, for field reconstruction purposes. While data-driven methods are available, their generalizability across different flow conditions (\textit{e.g.,} different Reynold numbers) remains questioned. Moreover, they frequently face the problem of spectral bias, which leads to smooth and blurry reconstructed fields, thereby decreasing the accuracy of reconstruction. We introduce FLRNet, a deep learning method for flow field reconstruction from sparse sensor measurements. FLRNet employs an variational autoencoder with Fourier feature layers and incorporates an extra perceptual loss term during training to learn a rich, low-dimensional latent representation of the flow field. The learned latent representation is then correlated to the sensor measurement using a fully connected (dense) network. We validated the reconstruction capability and the generalizability of FLRNet under various fluid flow conditions and sensor configurations, including different sensor counts and sensor layouts. Numerical experiments show that in all tested scenarios, FLRNet consistently outperformed other baselines, delivering the most accurate reconstructed flow field and being the most robust to noise.
\end{abstract}


\section{Introduction}
\label{sec:intro}

The ability to rapidly reconstruct coherent flow fields from limited sensor observations is essential in many areas of physical sciences and engineering, such as computational and experimental fluid dynamics, aerospace engineering, and climate science \cite{PhysRevFluids.4.103907, annurev-fluid-010816-060042, Wu2023, Caverly2022, Tan2023}. Mathematically, this problem is described as follows. Given a complete flow field $\mathbf{u}(t)$ that returns the value in $\mathbb{R}^m$ and the measurement operator $\mathcal{H}: \mathbb{R}^m \to \mathbb{R}^p$ that provides its punctual measurement $\mathbf{y}(t)$ that gives values in $\mathbb{R}^p$, the goal of the flow field reconstruction task is to find the forward map $\mathcal{G}: \mathbb{R}^p \to \mathbb{R}^m$, which is, in theory, the inverse of $\mathcal{H}$. However, finding the map $\mathcal{G}$ as the inverse of $\mathcal{H}$ is by no means straightforward. It is because the complete flow field $\mathbf{u}$ often has a higher dimension than its punctual measurement $\mathbf{y}$, $m \gg p$; therefore, the measurement operator $\mathcal{H}$ is likely ill-conditioned and thus not invertible \cite{DUBOIS2022110733}. As a result, directly deriving $\mathcal{G}$ by inverting $\mathcal{H}$ is nearly impossible.

Several approaches are available to overcome the aforementioned problems and reconstruct the flow field from limited sensor measurements. \textit{Direct reconstruction} \cite{Loiseau_Noack_Brunton_2018} refers to an approach that directly reconstructs the flow field via an optimization process. In this model-free method, the reconstructed fields are represented as a linear combination of reference modes, and optimization is used to derive their associated coefficients. \textit{Data assimilation} \cite{MONS2016255} is another approach that involves using a dynamic model to predict the flow field from its previous states and use sensor measurements to improve the forecast quality. Finally, \textit{regressive reconstruction}, the focus of this paper, is an approach that uses machine learning (ML) methods to directly learn the forward map $\mathcal{G}$ from data. Here, ML models are used to approximate the forward map, and the model parameters are derived by solving a training optimization problem in which the discrepancy between the machine-reconstructed field and the corresponding ground truth is minimized. Regressive reconstruction provides numerous computational efficiency benefits over direct reconstruction and data assimilation, as it allows for a single training and multiple reuses of the regression model with considerably faster computations.

In line with regressive reconstruction, there are several notable works that have successfully learned the forward map $ \mathcal{G} $ to reconstruct the flow field from sensor measurements. Erichson \textit{et al.} \cite{Erichson2020} proposed the use of a shallow network to approximate the forward map $ \mathcal{G} $ and train it using data. The authors have demonstrated an increase in reconstruction accuracy for the method compared to traditional direct reconstruction methods based on POD across different validation cases. Li \textit{et al.} \cite{li2024deep} investigated the use of different deep learning techniques for the reconstruction of high Reynolds number turbulent flow fields. Investigated models include a multi-layer perceptron (MLP) network, a convolutional neural network (CNN), and a generative adversarial network (GAN). The results show that while GAN is better at reconstructing small-scale vortex structures, MLP shows better performance with time-averaged flow fields and also yields a significant advantage in computational efficiency. Several other significant studies in this direction have been conducted by Wu \textit{et al.} \cite{Wu2023}, Zhao \textit{et al.} \cite{ZHAO2024108619}, Xie \textit{et al.} \cite{Xie2024}. Despite certain successes, the direct modeling approach still contains several limitations, such as poor generalization due to overfitting and convergence issues caused by vanishing/exploding gradients. Despite several proposed solutions to these issues such as the use of regularization methods such as L1/L2 regularization and batch normalization \cite{Erichson2020}, they still persist.

For this reason, the use of dimensionality reduction has been proposed to reduce the difference in the input and output dimensions of the problem to promote a better learning of the forward map $ \mathcal{G} $. In this approach, one first aims to find an efficient, low-dimensional representation $z$ of the flow field $\mathbf{u}$. Consequently, ML is used to learn the map $\mathcal{G}_z$ between the sensor measurements $\mathbf{y}_c$ and the learned low-dimensional representation $z$, which, in theory, is claimed to be easier than the direct map $ \mathcal{G} $. For instance, Dubois \textit{et al.} \cite{DUBOIS2022110733} investigated the use of several dimensionality reduction techniques, including both linear and nonlinear ones, to learn the low-dimensional representation of the flow field. The authors showed that by reducing the dimension of the output using latent representations and selecting the proper regression model, one can obtain an efficient reconstruction model that can accurately and rapidly reconstruct the flow field.

Despite the potential of the dimensionality reduction approach, most methods validate only with a single flow condition (e.g., a single Reynold number) \cite{DUBOIS2022110733, ZHAO2024108619, Erichson2020}, leaving their generalizability across different flow conditions unknown. In addition, one also needs to overcome another critical problem in field-based representation learning, namely \textit{spectral bias}. This is the phenomenon in which the ML models are biased toward global flow features with low frequency while commonly missing the local ones with high frequency. This issue causes the reconstructed field to smooth out, thus reducing its accuracy. In fact, this is a common problem in ML for physical sciences and engineering but has not been properly addressed \cite{WANG2021113938,Nguyen2024PARCv2PR}.

\textbf{Main contributions.}
In this work, we introduce a deep learning method, namely FLRNet, for flow field reconstruction from sensor measurements that can accommodate the above issues. FLRNet consists of two deep learning models: a Fourier-feature-based variational autoencoder (VAE) trained with added perceptual loss to learn the low-dimensional representations of flow fields and an fully connected (dense) network to correlate the learned representation with the corresponding sensor measurements. During inference, the fully connected network will derive the latent representation of the flow field given its sensor measurements, and the trained decoder will reconstruct the flow field from the derived latent representation.

The main contribution of FLRNet is twofold. \emph{First}, we introduce Fourier feature layers that help mitigate the spectral bias problem, enhancing the accuracy of the model reconstruction compared to other baselines, as will be shown in Section~\ref{sec:Results}.
\emph{Second}, we trained and tested FLRNet using two benchmark problems with various flow conditions (e.g., different Reynolds numbers) to validate its generalizability across different domains and operating conditions. As previously mentioned, this aspect is crucial for ensuring the practicality of the method, yet it has not received sufficient attention in prior work.

The remainder of the paper is structured as follows.
Section~\ref{sec:related-work} presents some related work.
Section~\ref{sec:Methodology} introduces the proposed method, including its architecture design and training mechanism. Section~\ref{sec:Results} presents the validation of our method with a legacy benchmark problem that is the flow around a circular obstacle. We also assess the effects of several elements on the accuracy of the reconstructed flow field, including the number of used sensors and the layout of the sensor placement. In Section~\ref{sec:Conclusion}, we discuss the advantages and disadvantages of the proposed method compared to other baseline methods and suggest several directions for improvement.

\section{Related Work}
\label{sec:related-work}

\subsection{Flow Field Reconstruction From Sensor Measurements}

There are three common approaches to reconstruct the flow field from limited sensor measurements \cite{DUBOIS2022110733}. \textit{Direct Reconstruction} \cite{Loiseau_Noack_Brunton_2018}, refers to an approach that directly reconstruct the flow field via an optimization process. This is a model-free method in which reconstructed fields are represented as a linear combination of reference modes, $ \mathbf{u} \approx \mathbf{\hat{u}} = \sum_{j = 1}^k \phi_j \nu_j, $ where $\phi_j$ and $\nu_j$ ($j \in [1,k]$) are respectively the reference modes and the associated coefficients. Here, modes $ \phi_j $ can be approximated using modal decomposition such as POD \cite{Taira20174013} or sparse representation \cite{103907}. The goal of the direct reconstruction approach is to find the coefficients vector $ \nu_j $ by minimizing the discrepancy between the reconstructed and the ground truth flow field, $ \mathbf{\nu} = \mathrm{argmin}_{\mathbf{\nu}} \Vert \mathcal{H}(\mathbf{u}) - \mathcal{H}(\sum_{j = 1}^k \phi_j \nu_j) \Vert_2^2$. 

The second approach, namely \textit{Data Assimilation}, is the another method that is often used for field reconstruction from sensor measurement. This approach involves the use of a dynamical model to estimate the field. Here the sensor measurements are input to improve the quality of the forecast. The dynamical model can be reduced-order model \cite{Noack2003335, pnas.1517384113} or deep learning-based model \cite{CHENG2024104877,Nguyen2024PARCv2PR,sciadv.add6868}. The drawback of such method is that the accuracy of reconstructed fields heavily depends on the quality of dynamical model which is typically hard to achieve. As the field evolved temporally, the accuracy of the dynamical model will be dropped as the result of error accumulation. 

Finally, \textit{Regressive Reconstruction} \cite{ARNAULT2016436,Erichson2020} is the third approach that uses ML methods to directly learn the forward map $ \mathcal{G}$ from data instead of solving an optimization as in direct reconstruction. Here, the ML model is used to approximate the map $ \mathcal{G}$, such $ \mathcal{G} (\mathbf{y}|\theta) $ with $ \mathbf{\theta} $ are model parameters. The training goal here is to find the optimal set of $ \mathbf{\theta} $ that minimize the discrepancy between the machine-reconstructed field and the corresponding ground truth, $ \mathbf{\theta} = \underset{\mathbf{\theta}}{\operatorname{argmin}} \left\Vert \mathbf{u} - \mathcal{G} (\mathbf{y}|\theta) \right\Vert_2^2$. Regressive reconstruction poses many computational efficiency advantages compared to direct reconstruction as the regression model can be trained once and re-used multiple times. In this work, we focus on regressive reconstruction due to its computational advantages compared to other methods.

\subsection{Deep Learning for Regressive Reconstruction}
Recently, deep learning has found a widespread application in field reconstruction from limited sensor measurements. Erichson \textit{et al.} \cite{Erichson2020} are among the first to propose the use of deep learning for regressive reconstruction. The authors employed a model known as the shallow network to directly learn the forward map from the observational space to the state space. The shallow network demonstrated an enhancement in reconstruction accuracy compared to the traditional direct reconstruction methods including both standard improved POD-based method. Peng \textit{et al.} \cite{PENG2023108539} proposed a hybrid deep learning framework that combines traditional mode decomposition with residual learning using MLP networks to enhance the accuracy of field reconstruction from limited sensor measurement of unsteady periodic flow fields.

Despite the early success of deep learning, the big difference in the dimension of the observational and state spaces still poses significant challenges for deep learning, often causing overfitting and reducing the generalizability of the model. For this reason, Dubois \textit{et al.} \cite{DUBOIS2022110733} proposed the use of dimension reduction techniques to resolve that big difference in the two spaces. The authors investigated the use of different dimensionality reduction techniques, including both linear and nonlinear ones. The results showed that the deep variational autoencoder yields the highest reconstruction accuracy and is the most robust to noise included in the sensor measurement. 

Beside the use of MLP networks, researchers have also pursued the use of convolutional neural networks (CNN) for field reconstruction from sensor measurements. Building upon the success in computer vision, many efforts have been conducted to apply CNN for field reconstruction. For instance, Liu \textit{et al.}  \cite{LIU2021120684} combined residual CNN with MLP to reconstructed the flow field from several measurable information. In another work, Peng \textit{et al.} \cite{PENG2022107802} proposed the use of U-net for temperature field reconstruction in heat sink designs. The authors represented the value and position of sensors using a two-dimensional map, transforming the field reconstruction problem into an image-to-image translation problem, which was then solved using U-net. The use of neural operators, with the discritization-invariant characteristic, is also another line of potential research that has just been explored. As demonstrated by Zhao \textit{et al.} \cite{ZHAO2024108619}, neural operators can provide zero-shot super resolution while achieving similar or even better reconstruction accuracy compared to traditional CNN-based methods. This implies that one can now train neural operators on low-resolution data, while still making inferences on high-resolution data. This capability of neural operators helps reduce the size of required training data, which is extremely valuable given the complexity of collecting scientific data.

Despite certain successes of earlier research, as mentioned above, the majority of those methods were only validated for a limited range of flow conditions, and retraining is required if a new flow condition is required. Secondly, as with the majority of field prediction problems in ML for science, spectral bias is still a big concern for most methods, as it reduces the accuracy of the reconstructed field both visually and quantitatively. The literature frequently reports that predicted fields tend to smooth out, particularly at the boundaries between phases \cite{Nguyen2024PARCv2PR,sciadv.add6868}. In this work, we aim to overcome these aforementioned issues.



\section{Methodology}
\label{sec:Methodology}

\subsection{Problem formulation and architecture Design}
\begin{figure*}[tb!]
    \centering
    \begin{subfigure}{\textwidth}
        \centering
        \includegraphics[width=.6\textwidth]{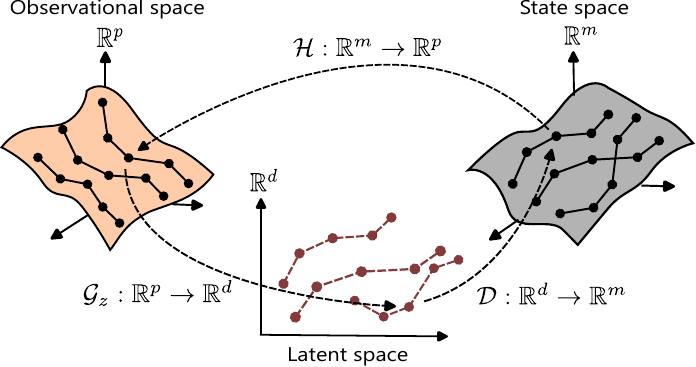}
        \caption{}
        \label{fig:problem_formulation}
    \end{subfigure}\\[0.3cm]
    \begin{subfigure}{\textwidth}
        \centering
        \includegraphics[width=0.75\textwidth]{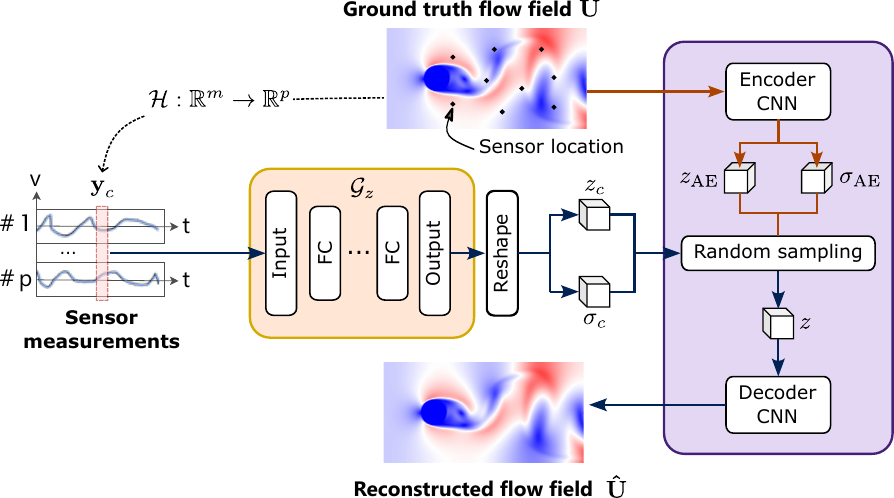}
        \caption{}
        \label{fig:overall-architecture}
    \end{subfigure}
\caption{The overall architecture design.}
\end{figure*}

Fig.~\ref{fig:problem_formulation} illustrates our problem formulation. In a finite-dimensional space, given the a snapshot of the flow field $\mathbf{u} \in \mathbb{R}^m$ at a given time $t$ and the map $\mathcal{H}: \mathbb{R}^m \to \mathbb{R}^p$ that provides the punctual measurement $\mathbf{y}$ of  $\mathbf{u}$, our goal is to find the find the forward map $\mathcal{G}: \mathbb{R}^p \to \mathbb{R}^m$ to reconstruct $\mathbf{u}$ from $\mathbf{y}$:
\begin{equation}
    \mathbf{\hat{u}} = \mathcal{G}(\mathbf{y}) \approx \mathcal{H}^{-1}(\mathbf{y})
\label{eqn:reconstruction}
\end{equation}
where $ \mathbf{\hat{u}}$ is the reconstructed flow field. 

As the dimension of $\mathbf{u}$ is much higher than that of $\mathbf{y}$, attempting to directly learn $\mathcal{G}$ from data can be complicated and is prone to overfitting \cite{DUBOIS2022110733}, thus, reducing the generalizability of the model. Therefore, we first aim to find the low-dimensional latent representation $\mathbf{z}$ of $\mathbf{u}$ to reduce the dimensional difference of the map between the observational and the state spaces. This representation learning task is accomplished by employing a variational autoencoder (VAE). 

VAE is a bottleneck-like neural network architecture consisting of two networks, namely, the encoder and the decoder. In VAE, the encoder $ \mathcal{E}$ learns to encode the flow field $\mathbf{u}$ into a latent representation $\mathbf{z}$; meanwhile, the decoder $ \mathcal{D}$ learns to accurately reconstruct the corresponding  flow field $\mathbf{u}$ from a given latent representation $\mathbf{z}$:
\begin{equation}
    \mathbf{\hat{u}} = \mathcal{D}(\mathbf{z}) =  \mathcal{D} \circ \mathcal{E}(\mathbf{u}).
\label{eqn:ae}
\end{equation}

The probablisitic learning strategy of VAE poses a few advantages. First, this learning strategy forces both the encoder $\mathcal{E}$ and decoder $\mathcal{D}$ to develop the capability to learn strong and robust features that effectively represent the ground truth full-state data $\mathbf{\hat{u}}$, enhancing the accuracy of the reconstruction process. Another well-known benefit of VAEs is that their latent space is more continuous and well-distributed compared to conventional autoencoders (AEs) \cite{Goodfellow-et-al-2016}. A common issue with AEs, particularly deep AEs, is the cluttered latent space that results from their attempts to overfit the training data. On the other hand, a well-trained VAE typically distributes its latent space evenly, avoiding significant gaps between data clusters. Therefore, the latent space of well-trained VAEs is more continuous than AEs, ensuring a well-distributed distribution of latent variables \cite{abs-1906-02691}. Such a smooth and well-distributed latent space will simplify the representation learning process for dynamical systems, as it minimizes significant jumps in the latent space traversal during the dynamic process.

After successfully learn the representation of the targeted flow field, we attempt to learn the map that correlate the sensor measurements $\mathbf{y}$ to the learned low-dimensional latent representation $\mathbf{z}$ of $\mathbf{u}$:
\begin{equation}
    \mathbf{z} = \mathcal{G}_z(\mathbf{y}).
\label{eqn:sens_latent_map}
\end{equation}

Finally, the reconstructed $\mathbf{\hat{u}}$ will be derived from the computed $\mathbf{z}$ via the learned decoding process:
\begin{equation}
    \mathbf{\hat{u}} = \mathcal{D} \circ \mathcal{G}_z(\mathbf{y})
\label{eqn:dim_reduced_recon}
\end{equation}
where $\mathcal{G}_z$ is the map from $\mathbf{y}$ to $\mathbf{z}$ and $\mathcal{D}$ is the map from $\mathbf{z}$ to $\mathbf{u}$, which is the decoding process. Following the regressive reconstruction method and inspired by the universal approximation theorem, we model $\mathcal{E}, \mathcal{D},$ and $\mathcal{G}_z$ using neural networks and attempt to to learn them from data. 

Fig.~\ref{fig:overall-architecture} illustrates the overall architecture design, which is the realization of the mathematical formulation described in~\cref{eqn:reconstruction,eqn:ae,eqn:sens_latent_map,eqn:dim_reduced_recon}. The proposed deep learning architecture is divided into two parts. In the first part (purple block), a variational autoencoder (VAE) is employed to learn a low-dimensional but rich representation of the original flow field. In the second part (orange block), a multilayer perceptron (MLP) network is used to correlate the sensor data with the latent representation of the corresponding flow field. For the VAE, we use a Fourier-feature-based, fully convolutional architecture with an additional loss term besides the conventional VAE loss, namely the perceptual loss. These added extra features to the neural network design and training will enable the VAE to learn perceptually rich features that are aware of the dynamic characteristics of flow fields, enhancing its reconstruction accuracy, as will be shown in the result sections. At the inference, the encoder $ \mathcal{E}$ is discarded, and only the decoder $ \mathcal{D}$ and the map $ \mathcal{G}_z$ are kept for the reconstruction of the flow field from sparse sensor measurements.

\subsection{Variational Autoencoder for Learning the Latent Representation of the High-fidelity Flow Field}
\label{sec:ae}

The first important component of our framework is to learn the rich, low-dimensional latent representation of the flow field. In this study, we accomplish this task by utilizing VAE.

VAE \cite{abs-1906-02691} is a special type of AE which contains two networks: encoder and decoder. In VAE, the encoder $\mathcal{E} (\phi)$ maps the original full-state data $\mathbf{{u}}$ to a distribution $\mathbf{z} \sim q(\mathbf{z)}$, often a Gaussian distribution. Meanwhile, the role of the decoder $\mathcal{D}$ is to accurately reconstruct the full-state $\mathbf{u}$ field from the latent distribution that is output from the encoding process $\mathbf{z} \sim q_\phi(\mathbf{z)}$, making it a probabilistic process. The training of VAE is accomplished by solving an optimization problem in which the below loss function is minimized:
\begin{equation}
\begin{split}
    \mathcal{L}_{\text{VAE}}(\phi, \theta) = - \mathbb{E}_{\mathbf{z} \sim q_{\phi}(\mathbf{z}|\mathbf{{u}})}\left[\log p_{\theta}(\mathbf{\hat{u}}|\mathbf{z}) + d_{\text{KL}}(q_{\phi}(\mathbf{z}|\mathbf{{u}}) \Vert p_{\theta}(\mathbf{z}))\right]
\end{split}
\label{eqn:vae_train}
\end{equation}
where $\mathbb{E}_{\mathbf{z} \sim q_{\phi}(\mathbf{z}|\mathbf{{u}})}(\log p_{\theta}(\mathbf{\hat{u}}|\mathbf{z})$ is the reconstruction loss, which aims to guarantee the accuracy of the reconstructed field $\mathbf{\hat{u}} $ compared to the ground truth field $\mathbf{{u}}$. In practice, the mean squared error (MSE) is often used for the reconstruction loss. Additionally, $p_{\theta}(\mathbf{\hat{u}}|\mathbf{z})$ is the probability distribution of the reconstructed flow field, while $q_{\phi}(\mathbf{z}|\mathbf{{u}})$ is the probability distribution of the latent variable. Here, $\phi$ and $\theta$ are network parameters of the encoder $\mathcal{E}$ and the decoder $\mathcal{D}$, respectively.  Finally, the $d_{\text{KL}}$ is the Kullback-Leibler divergence that measures the distance between the latent distribution produced by the encoder $\mathcal{E}$, $q_{\phi}(\mathbf{z}|\mathbf{{u}})$, and the target latent distribution $p_\theta(\mathbf{z})$ that is, in this case, a normal distribution $\mathcal{N}(0,1)$.

Fig.~\ref{fig:overall-architecture} illustrates our VAE design. Here, using the reparameterization trick, the output of our encoder includes the mean and variance vectors, and the latent vector that is used as the input for the decoder will be derived by random sampling from its mean and variance. We also employed a fully convolutional architecture to enhance the spatial scalability of our VAE.

\paragraph{Fourier Feature}

Fourier feature mapping  \cite{NIPS2007_013a006f} is a common technique in deep learning for positional encoding in many applications. Literature showed that Fourier features can effectively address spectral bias \cite{tancik2020fourfeat}, improving the learning capabilities of physics-informed deep learning models \cite{WANG2021113938}. The effectiveness of Fourier features against the spectral bias issue stems from their ability to control the frequency falloff, which is the primary factor preventing the network from learning high-frequency features \cite{tancik2020fourfeat}. In our work, we extend the traditional Fourier feature for MLP networks \cite{tancik2020fourfeat} to make it fit with our fully convolutional architecture as described below.

As described in Fig.~\ref{fig:overall-architecture}, we input to the encoder, beside the original flow field, a position field $\mathbf{x}:= [\mathbf{x}_1, \mathbf{x_2}]. $. Here, $\mathbf{x_1}$ and $\mathbf{x}_2$ are the horizontal and vertical coordinates of the corresponding pixel. Note that $\mathbf{x_1}$ and $\mathbf{x}_2$ have the same dimension with the original flow field. Consequently, we apply a Gaussian Fourier mapping $\gamma$ pixel-wise:
\begin{equation}
\begin{split}
    \gamma(\mathbf{x}) = \left[ \cos(\mathbf{B}x_1), \sin(\mathbf{B}x_1), \cos(\mathbf{B}x_2), \sin(\mathbf{B}x_2) \right]
\end{split}
\label{eqn:fourier_feature}
\end{equation}
where $\mathbf{B} \in \mathbb{R}^{m\times1}$ is randomly sampled from the normal distribution $\mathcal{N}(0,\sigma^2)$. Here, the number of mapping features $m$  and the variance of the normal distribution $\sigma$ are the hyperparameters that are preset during the training. In our experiments, we set $m = 4$ and $\sigma = 5$. 

\paragraph{Perceptual Loss}

We also enhance the representation learning with VAEs by introducing perceptual loss into the training. Perceptual loss is a common loss function in the computer vision community to enable deep generative models to produce high-fidelity synthetic images \cite{Johnson2016}. The goal of the perceptual loss function is to guide the VAE to learn to reconstruct the target snapshot image with a high level of fidelity and low discrepancy in not only the pixel space but also in a feature space. The use of loss functions defined on the feature space will overcome the lack of `perceptual' similarity measurement of per-pixel space and allow the network to learn to extract stronger and more meaningful features. Furthermore, the incorporation of perceptual loss will reduce the appeal of low-frequency modes in neural network training, thereby directing convergence toward more precise solutions.

The perceptual discrepancy between the two flow fields can be computed as
\begin{equation}
\begin{split}
    \mathcal{L}_{\text{perceptual}} = \left\Vert \Phi(\hat{\mathbf{u}}) - \Phi(\mathbf{u})\right\Vert_2^2
\end{split}
\label{eqn:perceptual_discrepancy}
\end{equation}
where $\mathbf{u}$ is the reconstructed flow field, $\hat{\mathbf{u}}$ is its corresponding ground truth, and $\Phi(.)$ is the activation map at the final convolution layer of a pretrained deep neural network \cite{Johnson2016}. In this work, we use Inception-V3 \cite{SzegedyVISW15}, which was pretrained with the Imagenet dataset \cite{5206848}.

\subsection{Correlating the Flow Field Latent Representation With Sensor Measurements}

The second part of our framework is to learn the mapping function $\mathcal{G}_z$ that correlates the punctual measurements of the flow field $\mathbf{y}$ to its corresponding low-dimensional latent representation $\mathbf{z}$. Fig.~\ref{fig:overall-architecture} illustrates the architecture of the $\mathcal{G}_z$ network. The measurement vector of the flow field at a given time $\mathbf{y}(t)$ is used as the input for an MLP network that is composed of five fully connected (dense) layers with 128 neurons used for each layer. The MLP network's output vector has a similar size to the dimension of the latent space. We then reshape the output vector to make it compatible with the latent representation learned with VAE. We use two convolutional layers with kernel size 3 to derive the predicted mean and variance of the latent variable from this reshaped tensor. Similar to VAE, we randomly sample the latent variable based on its previously computed mean and variance to use as input for the decoding process. We reconstruct the flow field using the same VAE decoder described in Section \ref{sec:ae}.
 
The training of the sensor mapping network can also be viewed as an optimization process, using the following loss function:
\begin{equation}
\begin{split}
    \mathcal{L}_{\text{mapping}}(\gamma) =  - \mathbb{E}_{\mathbf{z} \sim \mathcal{G}_z(\mathbf{z}|\mathbf{y})}\left(\log p_{\theta}(\mathbf{\hat{u}}|\mathbf{z}) \right) + d_{\text{KL}}\left(\mathcal{G}_z(\mathbf{z}|\mathbf{y}) \Vert q_{\phi}(\mathbf{z}|\mathbf{{u}}) \right).
\end{split}
\label{eqn:sensor_recon_loss}
\end{equation}

Notice that the loss function in Eq. \ref{eqn:sensor_recon_loss} consists of two terms. The first term $\mathbb{E}_{\mathbf{z} \sim \mathcal{G}_z(\mathbf{z}|\mathbf{y})}\left(\log p_{\theta}(\mathbf{\hat{u}}|\mathbf{z}) \right) $ is added to guarantee the latent representation predicted by the mapping network can be used to accurately reconstruct the flow field. Meanwhile, the term $d_{\text{KL}}\left(\mathcal{G}_z(\mathbf{z}|\mathbf{y}) \Vert q_{\phi}(\mathbf{z}|\mathbf{{u}}\right) $ is employed with the aim to minimize the distance between the predicted distribution of the latent representation predicted by the mapping network, $\mathcal{G}_z (\mathbf{z}|\mathbf{y})$, and the one produced by the encoder $q_{\phi}(\mathbf{z}|\mathbf{u})$.

\subsection{Training}

Our deep learning framework was trained via a two-stage training strategy. In the first stage, only the VAE was trained using the loss functions described in \cref{eqn:vae_train,eqn:perceptual_discrepancy}. The goal of this stage is to learn a rich latent representation of the original flow fields. In the second stage, the mapping network $\mathcal{G}_z$ was trained to learn to correlate the sensor measurements to their corresponding latent representation of the flow field. Note that in the second stage, we freeze the trained parameters of both the encoder and decoder. The loss function defined in Eq. \ref{eqn:sensor_recon_loss} is used for the second stage of training. In both stages, Adam Optimizer with learning rate of $1 \times 10^{-5}$ is used to solve the training optimization problem.


\begin{figure*}[bt!]
     \centering
     \begin{subfigure}{.248\textwidth}
         \includegraphics[width=\textwidth]{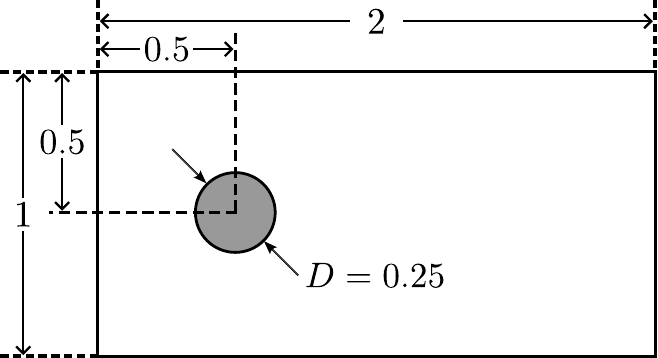}
         \caption{Domain's dimension.}
         \label{fig:domain_dim}
     \end{subfigure}
     \hspace{0.3cm}
     \begin{subfigure}{0.65\textwidth}
         \includegraphics[width=\textwidth]{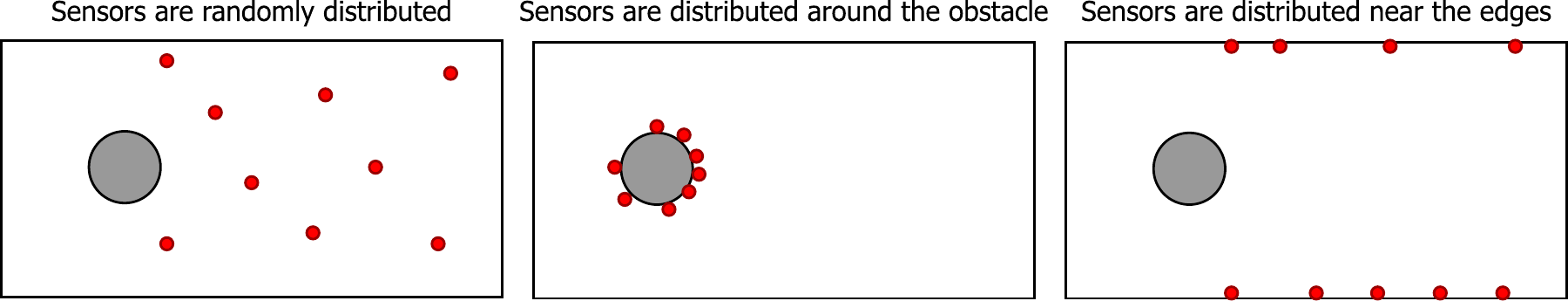}
         \caption{Three different tested sensor layouts.}
         \label{fig:setup_sensors_location}
     \end{subfigure}
     \caption{Numerical experiment setting for the flow around cylindrical test problem. (a) Dimension of the examined domain. (b) Three different tested sensor layouts.}
     \label{fig:exp-res}
\end{figure*}



\section{Results and Discussion}
\label{sec:Results}


We validate our method on the classic fluid flow behind a circular cylinder problem. In this problem, we examined fluid flow with Reynolds numbers ranging from $300$ to $1000$. The examined domain is a $1 \times 2$ (m) domain (see Fig.\ref{fig:domain_dim}), tessellated into a grid system with $128 \times 256$ grid points distributed uniformly. The cylindrical obstacle of size $ D = 0.25$ (m) with the center of the circle is placed at the coordinates $x = 0.5$ and $y = 0.5$ (m). In this problem, we focused on the velocity magnitude field of the fluid. This ﬂow field is characterized by a periodically shedding wake structure, having smooth and large-scale patterns. The numerical solution of the flow was obtained by solving the Navier-Stokes equation using the finite volume method. In this domain, we randomly place sensors to observe the flow velocity and aim to reconstruct the flow field from these sensor measurements. The sensor placement was finished using the Latin hypercube method. There are a total of $15$ simulation cases in the dataset, with $11$ used for training and four used for testing. In this experiment, we examined the reconstruction capability of our model in cases where $8$, $16$, and $32$ sensors were used. We also tested three different sensor layouts: sensors distributed randomly over the domain, sensors distributed around the cylindrical obstacle, and sensors distributed near the wall boundary of the domain (Fig.~\ref{fig:setup_sensors_location}). For the benchmarking purpose, we compare the field reconstruction of two FLRNet variants (with perceptual loss and with Fourier feature) with several baseline methods, including the multi-layered perceptron network (MLP), an upgrade version of the shallow network by Erichson \textit{et al.} \cite{Erichson2020} and the POD-based method \cite{DUBOIS2022110733}.

\subsection{Validation Results}

\begin{figure}[htb!]
     \centering
     \includegraphics[width= \textwidth]{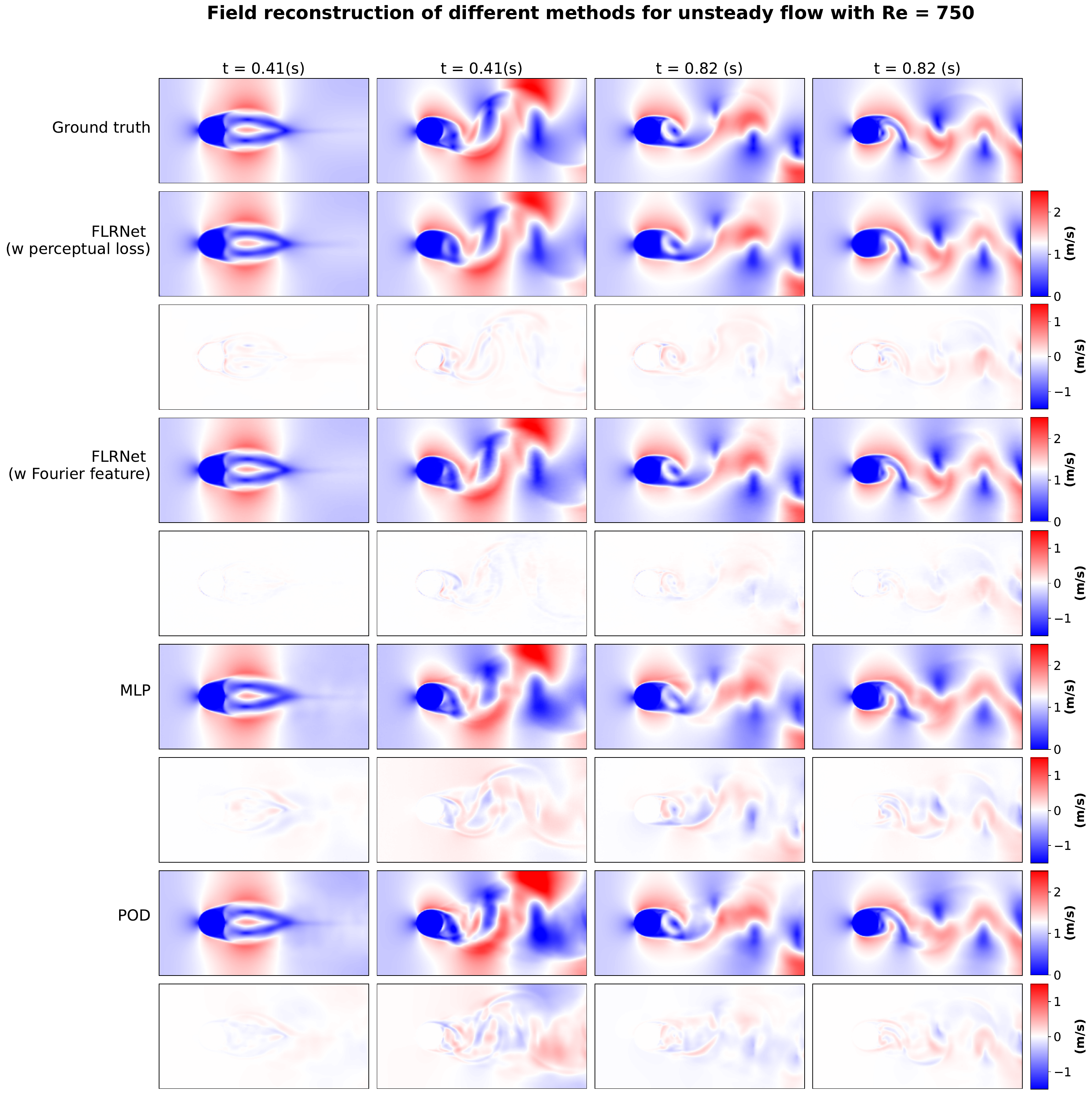}
\caption{Reconstruction result of FLRNet and other baselines at different times of the simulation. FLRNet reconstructed fields are closer to the ground truth compared to other baselines, indicated by the low-value error fields, especially at time $t = 0.82$ (s) and $1.23$ (s).}
\label{fig:flow_step_10}
\end{figure}

Fig.~\ref{fig:flow_step_10} shows the reconstructed fields by FLRNet compared to other baselines at different times of the simulation for the flow with $\text{Re} = 750$. Visually, all methods appear to be able to reconstruct the flow field from the limited sensor measurements, as the general pattern of the vortex shedding was recognized for all baselines. However, the flow fields reconstructed by FLRNet were the most accurate and closest to the ground truth derived by numerical simulation, as indicated by the low-value error fields with both configurations (with perceptual loss and with Fourier features). Meanwhile, the fields reconstructed by MLP and POD-based methods appear to be blurry and smooth. This can be clearly observed at the tail of the flow features at $t = 0.41(s)$ and at the boundary of the oscillatory vorticity feature at $t = 0.82$ and $1.23(s)$. This is because both methods, to some extent, rely on regularization (L1, L2 for MLP and the use of orthogonal bases in POD) to prevent overfitting of the training data. However, this regularization negatively impacts the quality of the field reconstruction, particularly at the feature boundary, leading to blurry and smooth reconstructed fields.

\begin{figure}[htb!]
     \centering
     \includegraphics[width=0.9\textwidth]{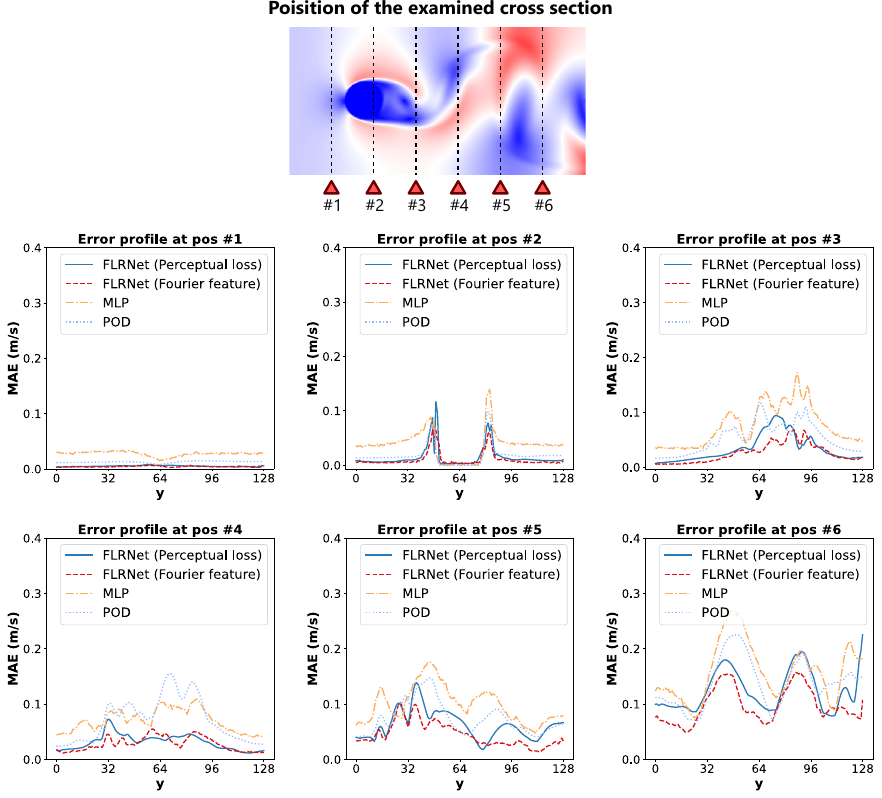}
\caption{Error profile analysis. We computed the average MAE across the whole test dataset at six different horizontal positions over the flow domain. Notice that the average MAE of FLRNet with Fourier feature is constantly the lowest for all examined positions. }
\label{fig:profile_analysis}
\end{figure}

We further investigated the performance of FLRNet and other baselines by analyzing the MAE profile at different vertical cross sections. Fig.~\ref{fig:profile_analysis} shows the average MAE at six different horizontal positions as indicated. From Fig.~\ref{fig:profile_analysis}, FLRNet has a lower MAE compared to other baselines (for all configurations). Interestingly, the plot in Fig.~\ref{fig:profile_analysis} also supports some of our hypotheses. First of all, it can be seen that MLP has the highest MAE for all examined sections. This phenomenon holds true for all other reconstruction results. It indicates that the use of dimensionality reduction can help enhance the reconstruction accuracy even with a simple linear model like POD. Furthermore, FLRNet, utilizing VAE, a non-linear dimension reduction technique, demonstrates superior reconstruction accuracy when compared to linear methods like POD. This result, again, confirm the advantage of using nonlinar dimension reduction methods as opposed to linear ones. Finally, FLRNet with Fourier features seems to produce the reconstructed flow field with the velocity magnitude closer to the ground truth compared to FLRNet with perceptual loss. This indicates that the use of Fourier features is more effective in battling the spectral bias issue compared to the use of perceptual loss. Those above observations from Fig.~\ref{fig:profile_analysis} can be confirmed by the quantitative validation using root mean squared error (RMSE) reported in Table~\ref{tab:pred_acc}, where FLRNet with Fourier features shows the lowest MAE while MLP hits the highest values.

\begin{table*}[htb]
\caption{Average reconstruction accuracy of FLRNet and the other baseline models.}
\label{tab:pred_acc}
\vskip 0.15in
\begin{center}
\begin{small}
\begin{sc}
\begin{tabular}{lccccc}
    \toprule
    & \multicolumn{5}{c} {Velocity mean absolute error (m/s)} \\
        
    Model  & \multicolumn{3}{c} {Random distribution} & {Around obstacle} & {Near edge} \\
           & $8$ sensors & $16$ sensors & $32$ sensors & $32$ sensors & $32$ sensors \\
    \midrule
    FLRNet (Percep. Loss ) & $0.028$ & $0.026$  & $0.021$  & $0.038$  & $0.023$  \\
    FLRNet (Fourier Feat.) & $0.022$ & $0.018$  & $0.016$  & $0.033$  & $0.018$  \\
    MLP                    & $0.056$ & $0.055$  & $0.046$  & $0.065$  & $0.048$  \\
    POD                    & $0.044$ & $0.037$  & $0.036$  & $0.052$  & $0.042$  \\
    \bottomrule
\end{tabular}
\end{sc}
\end{small}
\end{center}
\vskip -0.1in
\end{table*}

\subsection{Temporal Analysis}
\begin{figure}[tb!]
     \centering
     \includegraphics[width=0.6\textwidth]{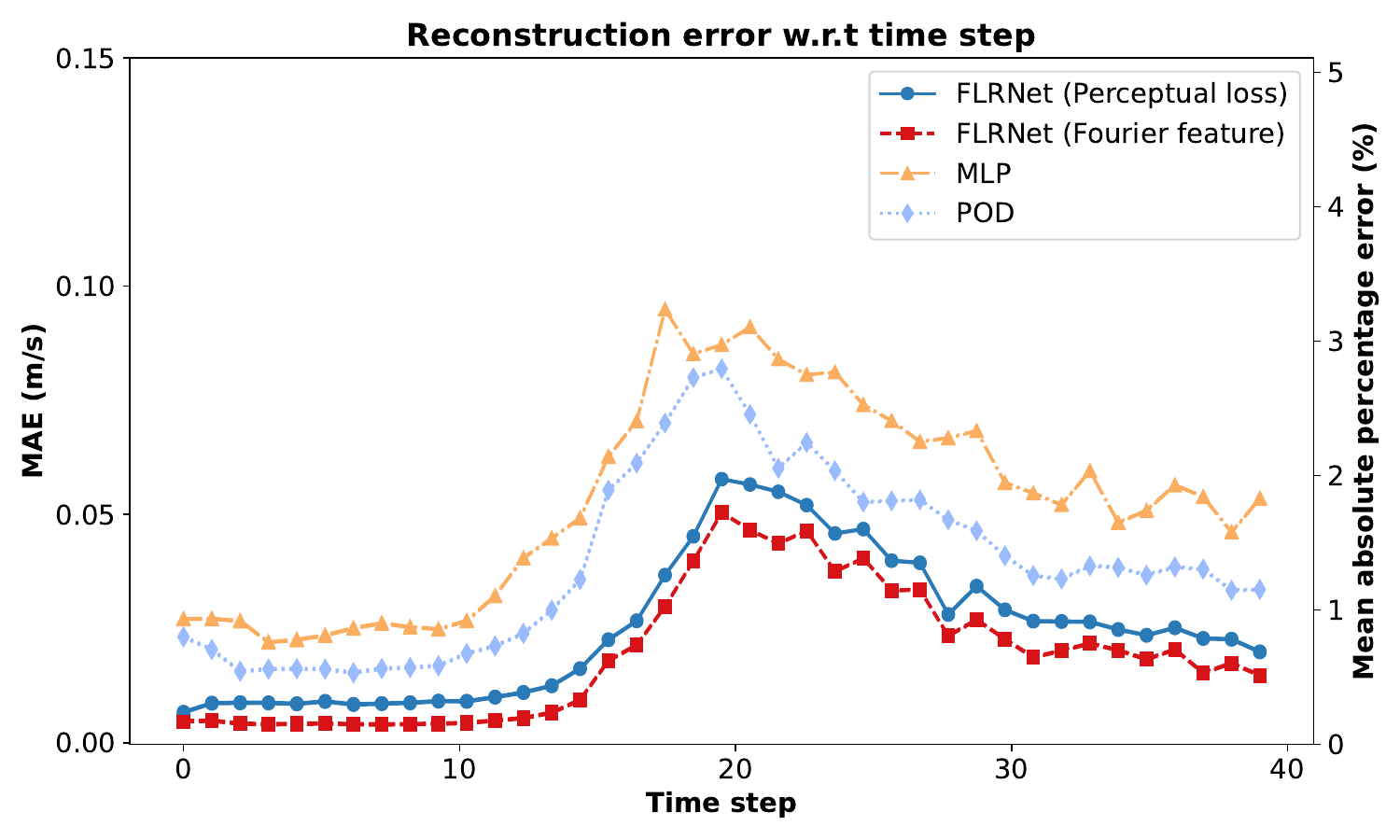}
\caption{Analysis of the reconstruction error w.r.t the temporal evolution of the flow field. The reconstruction error increases for all models, peaks at around step 20 when the instability arises and gradually reduces as the flow enters its equilibrium. During the whole process, FLRNet with Fourier feature yield the lowest MAE among all baselines.}
\label{fig:temporal_analysis}
\end{figure}

We also analyze the temporal evolution of the reconstruction error of FLRNet to investigate the model performance at different stages of the simulation. In this experiment, we measure the variation in the average MAE of ML reconstructed fields as the simulation progresses. Fig.~\ref{fig:temporal_analysis} depicts the variation in the MAE of the predictions by FLRNet and other baselines as the simulation time increases. As shown in Fig.~\ref{fig:temporal_analysis}, the MAE of all models remains stable during the first quarter of the simulation. This is the stage of the simulation where the fluid enters the domain and the flow field is still in a stable state. The MAE value then increases drastically and peaks at around step $20$, which is equivalent to around $1$ (s) in the simulation. This is when the flow instability arises and vortex shredding begins. Across the whole simulation, the period when flow instability arises is the most chaotic and most difficult to capture, which explains the peak in MAE value for all models. After reaching the peak, the MAE of all models gradually decreases. This is coincident with the stage where the flow enters its equilibrium and starts forming the vortex pattern. Compared to the previous stage of the flow, the final stage is more straightforward and easier to capture. Also, notice that throughout this process, the MAE of FLRNet is consistently the lowest among all other baselines, demonstrating its better reconstruction capability compared to other baselines during all stages of the simulation.

\begin{figure*}[tb!]
     \centering
     \begin{subfigure}{.4\textwidth}
         \includegraphics[width=\textwidth]{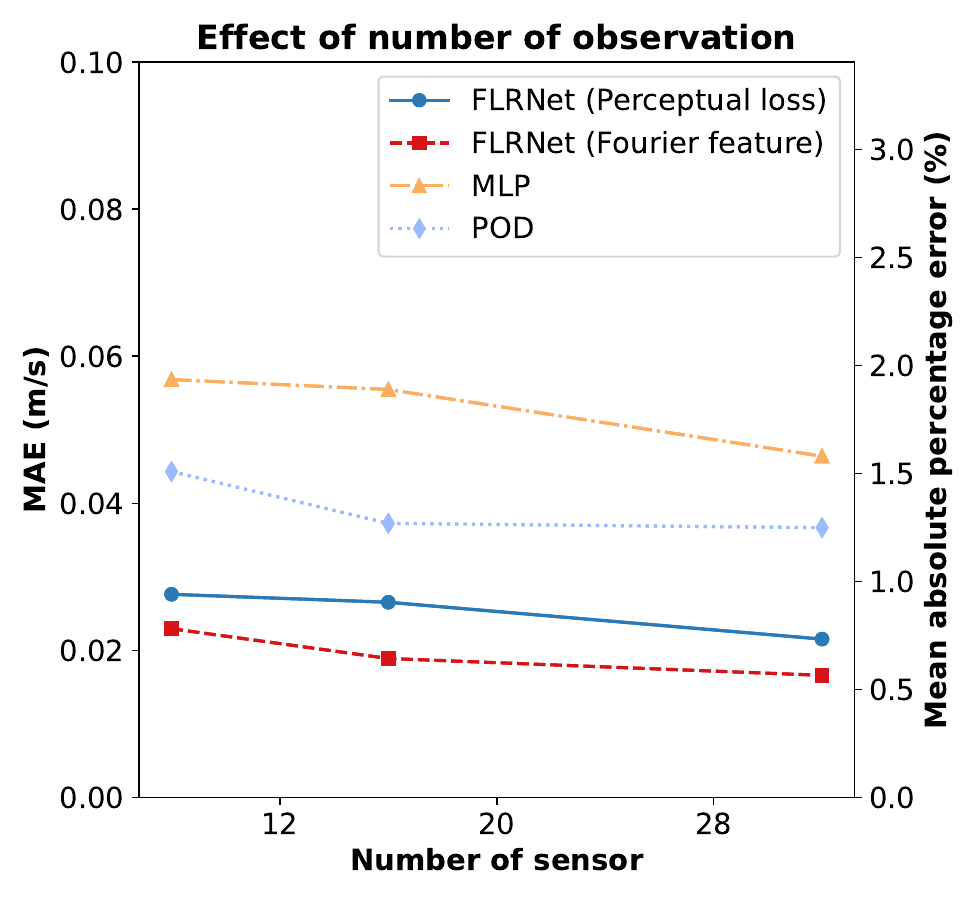}
         \caption{}
         \label{fig:no_of_sensor}
     \end{subfigure}
     \begin{subfigure}{.4\textwidth}
         \centering
         \includegraphics[width=\textwidth]{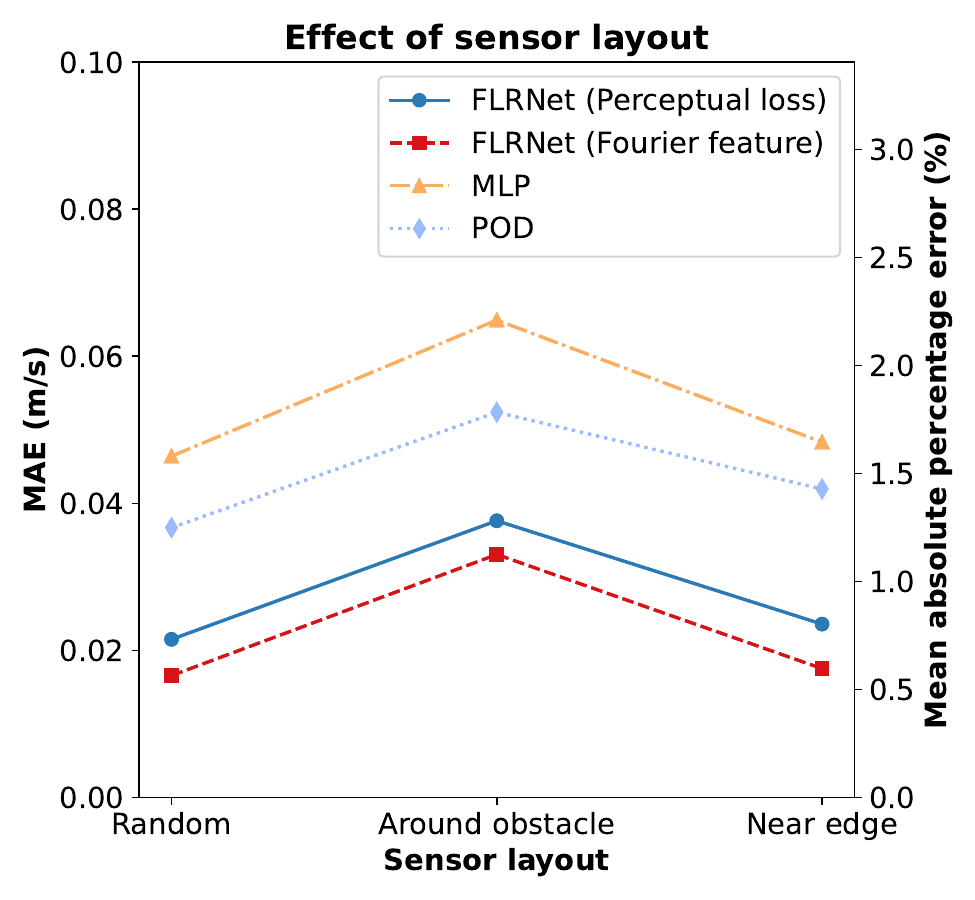}
         \caption{}
         \label{fig:sensor_layout}
     \end{subfigure}
     \caption{Effect of different sensor configurations on the performance of FLRNet and other baselines. (a) Effect of number of sensors used. (b) Effect of sensor layout.}
     \label{fig:sensor_effect}
\end{figure*}

\subsection{Effect of Sensor Placements and Sensor Counts}

We continue to investigate the effect of different sensor settings on the performance of FLRNet, including the number of sensors used and the layout of the sensor distribution. For the number of sensors, we varied the number of used sensors from $8$, $16$, and $32$ sensors. For sensor layout, we tested three different layout settings: randomly distributed, positioned around the cylinder, and distributed near the boundary of the domain. Finally, we compute the average MAE for each sensor configuration and compare them to assess the effect of different sensor settings.

Fig.~\ref{fig:no_of_sensor} illustrates how the number of sensors affects the accuracy of all models. From Fig.~\ref{fig:no_of_sensor}, it can be concluded that an increase in the number of sensors leads to an increase in reconstruction accuracy across all baselines. This can be explained by the fact that the number of sensors increases, which leads to more information about the full-state flow field that can be captured. As a result, it becomes easier for the ML models to learn the characteristics of different snapshots of the flow field and accurately reconstruct them from the sensor measurement.

\begin{figure}[htb!]
     \centering
     \includegraphics[width=0.4\textwidth]{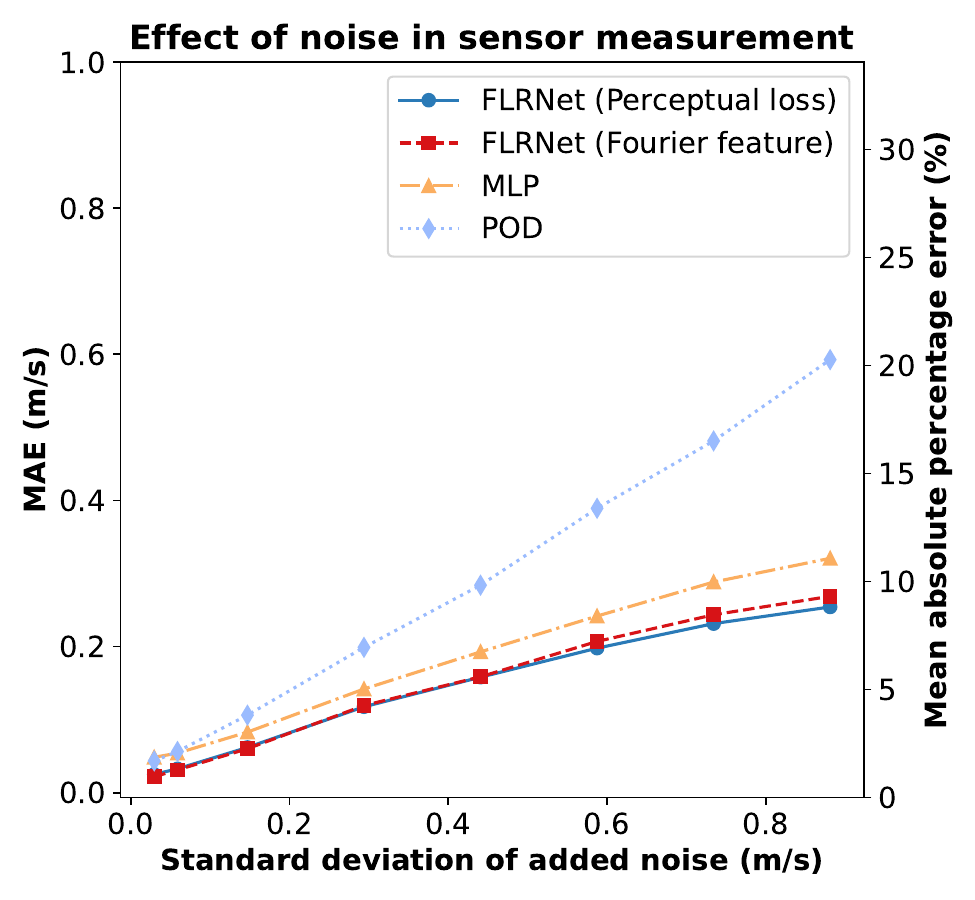}
\caption{Effect of noise in sensor measurement. Compared to other baselines, FLRNet is the most robust method as its MAE is the lowest as the level of noise increase. }
\label{fig:noise_effect}
\end{figure}

\begin{figure}[htb]
     \centering
     \includegraphics[width=0.4\textwidth]{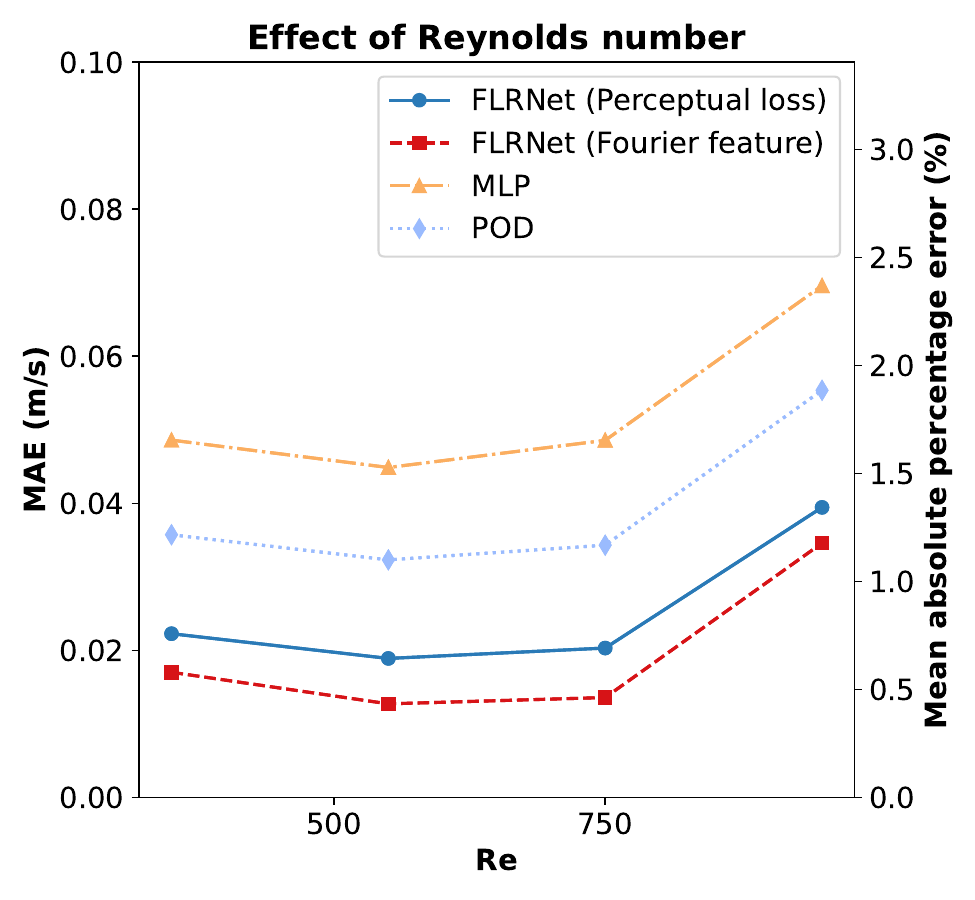}
\caption{Reconstruction error of FLRNet compared to other baselines for different flow conditions.}
\label{fig:reynold_number_effect}
\end{figure}

Fig.~\ref{fig:sensor_layout} illustrates the impact of sensor layout when $32$ sensors are used. Here, we exclusively examine the case of $32$ sensors, as our preliminary experiment indicates that this configuration yields the best results for all baseline models. From Fig.~\ref{fig:sensor_layout}, it is evident that the reconstruction accuracy in the case of a sensor randomly distributed over the domain is the highest. Meanwhile, the layout in which sensors are distributed around the cylindrical obstacle is the most difficult to reconstruct, as the reconstruction accuracy for all baselines is the lowest among the three layouts tested. This result is expected, as when the sensors are evenly distributed in the domain, information is captured for different areas that represent different facets of the flow fields. This information leads to a better characterization of the flow field, resulting in better reconstruction accuracy. Meanwhile, in the case of a sensor distributed around the cylindrical obstacle and near the boundary of the flow field, the information about only a single region (near the boundary) of the flow field is captured. Therefore, in this scenario, reconstructing the entire state of the flow field presents a significant challenge for the ML model.

Finally, the result showed in Fig.~\ref{fig:sensor_effect} that FLRNet consistently outperforms other baselines in all tested sensor configurations. Its MAE for all testing configurations is the lowest among all comparison baselines. Fig.~\ref{fig:sensor_effect} also demonstrates the robustness of FLRNet, as its MAE is still stable when the sensor layout varies. Even in the case of sensors distributed around cylindrical obstacles and near the edge, reconstructed fields by FLRNet have a lower MAE compared to ones by MLP and POD. These results demonstrate FLRNet's versatile reconstruction capability, making it applicable in numerous real-world application scenarios.

\subsection{Effect of Noise Addition}

FLRNet is trained on a simulation, clean dataset of fluid flow; however, the real-world sensor measurement in practice always contains noise. Therefore, we validate the robustness of FLRNet by assessing their prediction accuracy at different levels of noise added to the sensor measurement. In this experiment, we assume the sensor measurements of the flow field follow the Gaussian distribution:
\begin{equation}
\begin{split}
    \mathbf{y}_{\text{noise}} =  \mathbf{y} + \mathcal{N} (0, \sigma^2).
\end{split}
\label{eqn:noise_added}
\end{equation}

We vary the level of the added noise by varying the variance $\sigma^2$ and tested the responses of FLRNet and other baselines. Fig.~\ref{fig:noise_effect} shows the MAE of FLRNet and other baselines when the levels of added noise vary. From Fig.~\ref{fig:noise_effect}, it can be seen that when there is an appearance of noise in the sensor measurement, the reconstruction accuracy of all models reduces. Additionally, one can conclude that FLRNet is more robust to noise in the sensor measurement than MLP and POD, as the rate of accuracy drop when the noise level increases in FLRNet is significantly lower than in MLP and POD.

\subsection{Generalization Across Different Flow Conditions}

The generalizability of FLRNet across different flow conditions, particularly the Reynolds number, is another advantage besides its prediction capability. Fig.~\ref{fig:reynold_number_effect} illustrates the mean average error (MAE) of FLRNet and other baselines on the test set as the Reynolds number of the flow varies. Notably, as the Reynolds number of the flow approaches the training boundary, the MAE of FLRNet increases. In particular, we observed higher FLRNet errors at $\text{Re} = 350$ and $\text{Re} = 1000$, and lower ones at $\text{Re} = 550$ and $\text{Re} = 750$. This behavior of FLRNet can be explained by the fact that ML models tend to perform better when the test samples are within the training set, \textit{i.e.}, the model is performing the interpolation task. When the test samples reach or exceed the training boundaries, \textit{i.e.}, the model is performing an extrapolation task, its performance tends to decrease. This pattern is also consistent across all baselines. Despite the effect of the Reynolds number on the reconstruction accuracy, the MAE of FLRNet is consistently the lowest across all baselines.

\section{Conclusion}
\label{sec:Conclusion}

We present FLRNet, a deep learning method, for flow field reconstruction from limited sensor measurements. FLRNet employs a VAE network to learn a low-dimensional latent representation and an MLP network to correlate the sensor measurement to the learned representation. FLRNet equips its VAE with Fourier feature layers and adds perceptual loss to enhance its ability to learn detailed features of the flow field. Experimental studies showed that added Fourier feature layers and perceptual loss help FLRNet overcome the spectral bias issue, enhance its reconstruction accuracy, and increase its robustness to noise in sensor measurement compared to other traditional methods relying solely on MLP networks or linear dimension reduction techniques such as POD. Additionally, the results of the numerical experiment show that FLRNet exhibits high generalizability across various flow conditions when compared to other baselines, which is the current gap in the literature. Future works will expand the capability of FLRNet to cope with different sensor settings, including the variation in the number of used sensors and their positions. We are also working on enabling FLRNet to handle random discretization grid systems in both the spatial and temporal domains under investigation. This is a missed opportunity in the current design of FLRNet. Finally, we are preparing to expand the application of FLRNet for other domains, such as material science or experimental fluid dynamics.

\section*{Acknowledgments}

The authors acknowledge the support of Phenikaa University, Hanoi, Vietnam, during the conduct of this research.


\bibliographystyle{unsrt}  
\bibliography{references}

\end{document}